\documentstyle[epsfig]{article}
\begin{document}
\title{Regge Trajectories For All Flavors}
\author{Silvana Filipponi\footnote{Silvana@feynman.harvard.edu 
and Silvana@pg.infn.it}[1,2],Giulia
Pancheri\footnote{Pancheri@lnf.infn.it}[3] 
and Yogendra Srivastava\footnote{Srivastava@pg.infn.it}[2,4]}
\maketitle
\centerline{HUTP-97/A065} \vskip5mm
\centerline{1.{\em Physics Department, Harvard University, Cambridge, MASS, USA}}
\centerline{2.{\em Dipartimento di Fisica e INFN, Universita' di Perugia, 
Perugia,Italy}}
\centerline{3.{\em Laboratori Nazionali di Frascati dell' INFN, Frascati,
Rome, Italy}}
\centerline{4.{\em Physics Department, Northeastern University, Boston, MASS, USA}}
\begin{abstract}
Based on available data 
for mesonic resonances of light, medium and heavy
flavors, we have performed a global analysis to construct the corresponding
linear Regge trajectories. These have been supplemented by results from various
phenomenological models presented in the literature. A satisfactory formula is
found for the dependence of the intercept and the slope on quark masses. We
find reasonable agreement with data on production of charmed hadrons through
exchange of our charmed trajectories in the space-like region. When applied to
mesons containing the top quark, our results suggest the impossibility of their
formation as evidenced by other independent analyses.
\end{abstract}
\vskip18mm
Exactly how quarks are bound together to form hadronic bound states is not
known. QCD - the underlying theory for strong interactions - is flavor
independent and it has been found difficult to compute the dependence of
mesonic bound states (say) on the quark masses. In consequence, quite different
approximation schemes and models exist which try to obtain the mass spectra for
different flavors. These may be roughly classified as follows.\par
\noindent (i) light-light systems: mesons composed of u, d and s quarks ( 
$m_{quark}$ $< < \Lambda$) are described by ultrarelativistic 
Bag models or Chiral theories, where quarks are treated essentially as massless. 
\par
\noindent (ii) Charmonium and Bottomium systems: Since both quarks are heavy
($m_c$, $m_b$ $>> \Lambda$), non-relativistic models with a central
potential are invoked. A smooth function interpolating between the Coulomb term
and a confining linear term gives the short and the long distance behaviors.
Other potentials include a constant in addition to a small power in $r$.\par
\noindent (iii) light-heavy systems ($B$ and $D$ mesons) are poorly known
experimentally and phenomenologically: a heavy quark mass expansion has often
been employed.\par
It is fair to conclude that there is as yet no overall picture providing an
understanding of all the mesonic bound states.\par
In this work we describe a different approach based on the Regge
trajectories to answer the generic question posed above: how do mesonic
masses depend upon the quark masses. This formalism is particularly suited
for strongly interacting systems since (i) Regge trajectories exist for all
flavors, for large or small quark masses and (ii) angular momentum becomes a
continuous variable facilitating interpolation. Redundancy of data allows for
self consistency checks and the theoretical exercise is not merely curve
fitting. Morover, as we show later, the same Regge trajectories constructed
in the time-like region can be employed in the space-like region 
(exchange processes) as predictive tools. Our results a posteriori confirm this
endeavor.\par
Based on much trial and experimentation, we made the following two crucial
simplifying assumptions to obtain the Regge trajectories: \par
\noindent (i) All trajectories were assumed {\bf linear} in $(mass)^2$ of the
hadronic state,
$$
\alpha(s) = \alpha(0) + s \alpha' \eqno(1)
$$
so that a meson of angular momentum $J$ has the $(mass)^2$ equal to $s_J$. For
light mesons (and baryons), we {\it know} this to be true experimentally.\par
\noindent (ii) The functional dependence of the two parameters $\alpha(0)$
and $\alpha'$
on the quark masses is via ($m_1\ +\ m_2$).\par
	For light mesons, experimental data [1] and some  
models [2] give the value, for the light-light system
$$
\alpha' \approx 0.9\ GeV^{-2} \eqno(2)
$$
We supplement this with experimental data [1], theoretical models and
phenomenological spectra for charmonium and bottomium systems [3,4,5]  
to obtain the following simple formula giving a global description
$$
\alpha'(m_1 + m_2) = {{0.9\: GeV^{-2}}\over{[1 + 0.22\:({{m_1 +
m_2}\over{GeV}})^{3/2}]}}  \ \eqno(3) 
$$
where $m_1$ and $m_2$ are the corresponding quark masses for that trajectory.
We show in Fig.(1), how well (3) compares with available data and some
models.\par
An analysis similar to the above was performed for the intercept
$\alpha_{I}(0)$, where the subscript $I$ refers to the leading trajectory. In
this work we limit ourselves to those mesonic systems for which the lowest
physical state is at $J$ $=$ $1$.  A global description for these is given by 
$$
\alpha_{I}(m_1 + m_2 ; 0) = 0.57 - {{(m_1 + m_2)}\over{GeV}}\eqno(4)
$$
A comparison of Eq.(4) with input data is shown in Fig(2). We note that only
two points from a theoretical analysis [6] fall quite below our curve. These
points refer to the $B_c$ (composed of quarks $b$ and $c$) system.  \par
We next consider the secondary Regge trajectories and splitting between the
energy levels for same $J$ of a given system. Calling $\alpha_I(0)$ the leading
and $\alpha_{II}(0)$ the secondary Regge trajectory intercept, we estimated the
``distance''
$$
\Delta\alpha(0) = \alpha_I(0) - \alpha_{II}(0) \eqno(5)
$$
from data (not very precise) and phenomenological models. We find a rather
loose bound\footnote{Leading trajectories whose ground states begin at $J$ $=$
$0$ (corresponding to the pion, or $\eta_c(1S)$, etc.), which have an
intercept approximately $0.5$ lower from the leading trajectory, follow a
similar pattern, but they are not considered here.}
$$
1.3 < \Delta\alpha(0) < 1.6 .\eqno(6)
$$
The result becomes more interesting when transposed in terms of a physically
more amenable quantity, viz., the energy splitting 
$$
\Delta E_J^{II - I} = (E_J^{II} - E_J^I)  \eqno(7)
$$
between the states of a given angular momentum of a given system. Quite
strikingly, it is found to be a constant (between $0.5 - 0.8$ GeV) for all
systems (composed of u, d, s; c and b quarks). It is shown in Fig(3). 
The approximate constancy of this energy difference for all $J$ and all
flavors, gives us confidence 
in the generality of this result and we expect it to be verified in all viable 
phenomenological models.  \par  
\par
Combining (3) and (4), we are in a position to give a general expression for
the leading mesonic Regge trajectory formed by any two quarks of masses $m_1$ 
and $m_2$ 
$$
\alpha (m_1 + m_2; t) = 0.57 - \frac{(m_1 + m_2)}{GeV} + 
{{0.9\:GeV^{-2}}\over{[1 + 0.22\:({{m_1 + m_2}\over{GeV}})^{3/2}]}}\;t  \eqno(8)
$$
In Fig(4), we show the leading Regge trajectories for different flavors for
space- and time- like regions ($- 500 GeV^2 < t < 500 GeV^2$). In this figure, we
have included the ``top'' and the ``toponium'' trajectories as well, even though,
as we shall demonstrate later, top and toponium bound states cease to exist as
physical states due to the fast weak decays of the top quark into a $W$ and a
$b$ quark.\par 
Eq.(8) allows us to make ``predictions'' (or consistency checks) about the
energy spectra of excited mesons for the $D$, $D_s$ and $B$, $B_s$ systems, which 
were not used as input data. In table I we show our results for the states of the
leading trajectory only. Whenever data exist, there is reasonable
agreement. A more complete spectrum for mesons may be found
in [7,8].\par   
As another application, we employ the analytic continuation into the
space-like  region of  these trajectories constructed through data in the
time-like (resonance) region, to discuss the inclusive
production of heavy flavours through the di- triple Regge formalism. Data exist
for $D$ [9,10,11] and $\Lambda_c$ [12] production reactions,
$$\pi + N \rightarrow D + X \ and \ \pi + N \rightarrow \Lambda_c + X  \eqno(9)$$
for a wide range of $x_F$. Assuming a factorized $p_t$ dependence, 
experimental data have been parametrized through
$$
{{d\sigma}\over{dx_F}} \approx (1 - x_F)^n  \eqno(10)
$$
with $n\approx (3.7 \div 3.9)$ [9,10,11] and $\approx 3.5$ [12].\par
We invoke the di-triple Regge formula [13,14], valid for large $M_X^2$
and  large  $(s/M_X^2)$,
$$
{{d^2\sigma}\over{dM_X^2 dt}} \rightarrow
{{\gamma(t)}\over{s}}\ \left ({{M_X^2}\over{s}}\right )^{1-2\alpha(t)} \eqno(11) 
$$
where $\alpha(t)$ is the exchanged Regge trajectory. Neglecting the
$t$-dependence, we have in this region 
$$
{{d\sigma}\over{dx_F}} \approx \left(1 - x_F\right)^{1-2\alpha(0)}  \eqno(12)
$$
For the reactions of Eq.(9), inserting a D- Regge trajectory exchange, we
have $\alpha(0)=-1.35$ (see Table I), leading to the exponent $n\ \approx\
3.7$, in satisfactory agreement with the data. We stress that
this is an important test since the space-like extrapolation depends crucially
on the slope parameter of the exchanged D trajectory (which is roughly a half
of that for the light system).  
\par
As a third application, let us consider the top system. It is generally believed
that the top quark cannot form bound states either with another $t$ quark or
with a light one. The point being that the top quark would decay into a real
$W$ and $b$ with a large width: the life-time would be even smaller than the
revolution time thereby precluding the formation of mesonic bound states 
[15].\par
Our analysis confirms the above physical picture quite nicely. The energy
splitting between the ground state and that lying on the second trajectory
are as follows. For the toponium, we find $1.3\ GeV\ <\ \Delta E_{Toponium}\ 
<\ 1.6\ GeV$, in good agreement with ref.[16]. For the top mesons, we find 
$1.1\ GeV\ < \Delta E_{Top}\ <\ 1.3\ GeV$. Neither can exist, since
$\Delta E_{Toponium}\ <\ 2\Gamma_t$ and $\Delta E_{Top}\ <\ \Gamma_t$.\par
In conclusion, we have obtained a global unified description of linear Regge
trajectories for all flavors, and their preliminary applications to space-
as well as time-like regions seem encouraging. Further work in many
directions is in progress. Inclusion of iso-spin,
charge and other quantum numbers can  be incorporated perturbatively. So far we
have also ignored the imaginary parts of the Regge trajectories $\alpha(s)$. A
knowledge of $Im\ \alpha(s)$ in the time-like region is useful since it is
directly related to the resonance widths. Through unitarity, it is also
connected to the residue function $\beta(s)$, which when continued to the
space-like region is relevant for the Regge exchange amplitudes. Another
interesting problem concerns the baryon Regge trajectories - a much more
difficult task since three masses are involved. If further work confirms the
viability of this approach, an effective interaction should be constructed
which is able to produce these results dynamically. We shall return to this
question elsewhere. Here we shall limit ourselves to making one general 
observation.
It appears that an interaction capable of generating linear Regge trajectories
requires a linear mass dependence in the confining term in order to reproduce
the $(m_1\ +\ m_2)^{-3/2}$ behavior in the slope.\par\noindent
\vspace{.5cm}\par\noindent
{\bf Acknowledgements}\par\noindent
S.F. would like to acknowledge the hospitality by Professor S. Glashow at
Harvard. G.P. and Y.S. would like to acknowledge some discussions with and much 
encouragement from Professor J.Bjorken.\par\noindent
\vskip15mm
\centerline {\bf REFERENCES:}\par
\vspace{.2cm}\noindent
1. Review of Particle Physics, Phys. Rev. D {\bf 54} (1996), 1-720.
\par\noindent
2. D. Flamm and F. Schoberl, Introduction to the quark model of elementary
particles, Vol.1 (Gordon and Breach Publishers, 1982); 
N.A.Tornqvist, in Stockholm 1990, Proceedings,{\it Low energy antiproton
physics}, p.287-303 and Helsinky University preprint HU-TFT-90-52.
\par\noindent
3. J. L. Richardson, Phys. Lett. B {\bf 82} (1979) 272.
\par\noindent
4. A. Martin, Phys. Lett. B {\bf 93} (1980) 338;  B {\bf 100} (1981) 511.
\par\noindent
5. A. K. Grant, J. L. Rosner and E. Rynes, Phys. Rev. D{\bf 47} (1993)
1981.
\par\noindent
6. J. Morishita, M. Kawaguchi and T. Morii,  Phys. Lett.  B {\bf 185}
(1987) 159;  Phys. Rev.  D {\bf 37} (1988) 159.
\par\noindent
7. S. Filipponi and Y. Srivastava, Invited Talk at Hadron97 
(BNL, N.Y., August 1997), ed. Suh-Urk Chung, to be published.
\par\noindent
8. S. Filipponi, G. Pancheri and Y. Srivastava, Invited Talk at 
XXVII Symposium on Multiparticle Dynamics (LNF Frascati,Italy, September
1997), to be published.
\par\noindent
9. E769 Collab., G. A. Alves  et al, Phys. Rev. Lett.{\bf 69} (1992) 3147; 
Phys. Rev. D {\bf 49} (1994) 4317. 
\par\noindent
10. LEBC-EHS Collab., M. Aguillar Benitez  {\it et al.}, Phys. Lett. {\bf 161B}  
(1985), 400.
\par\noindent
11. ACCMOR Collab., S. Berlag  {\it et al.}, Z. Phys. C {\bf 49} (1991), 555.
\par\noindent
12. ACCMOR Collab., S. Berlag  {\it et al.}, Phys. Lett. B {\bf 247} (1990) 113.
\par\noindent
13. A. H. Muller, Phys. Rev. D{\bf 2} (1970), 2963. 
\par\noindent
14. G. Pancheri and Y. Srivastava, Lett. Nuovo Cimento Vol.II (1971) 381.
\par\noindent
15. I. Bigi, Y. Dokshitzer, V. Khoze, J. Kuhn and P. Zerwas, Phys. Lett.
 {\bf B181} (1986) 157.
\par\noindent
16. N. Fabiano, A. Grau and G. Pancheri, Il Nuovo Cimento{\bf A107} (1994) 2789.
\vfill
\eject
\vskip15mm
Table I:{\it Leading trajectory parameters and predicted masses for D and
B mesons.}\par
\vskip10mm
\begin{center}
\begin{tabular}{|c|c|c|c|}\hline
\multicolumn{2}{|c|}{\bf $D$ mesons}&
\multicolumn{2}{|c|}{\bf $D_s$ mesons}\\
\hline \hline
\multicolumn{4}{|c|}{$\alpha '$ ($GeV^{-2}$ units)}\\ \hline
\multicolumn{2}{|c|}{0.586}&
\multicolumn{2}{|c|}{0.562}\\
\hline \hline
\multicolumn{4}{|c|}{$\alpha _I(0)$}\\ \hline
\multicolumn{2}{|c|}{-1.35}&
\multicolumn{2}{|c|}{-1.51}\\
\hline \hline 
\multicolumn{4}{|c|}{{\bf States of the Leading trajectories ($GeV$ units)}}\\ 
\hline 
{\it Predicted}&{\it Experimental}&{\it Predicted}&{\it Experimental}\\
\hline \hline
\multicolumn{4}{|c|}{J=1}\\ \hline
2.01&2.0067$\pm$0.0005
&2.11&2.1124$\pm$0.0007 \\
\hline
\multicolumn{4}{|c|}{J=2}\\ \hline
2.39&2.4589$\pm$0.0020&2.50&/ \\ 
\hline
\multicolumn{4}{|c|}{J=3}\\ \hline
2.73&/ &2.83&/ \\
\hline
\end{tabular}
\end{center}\par
\vskip5mm
\begin{center}
\begin{tabular}{|c|c|c|c|}\hline
\multicolumn{2}{|c|}{\bf $B$ mesons}&
\multicolumn{2}{|c|}{\bf $B_s$ mesons}\\
\hline \hline
\multicolumn{4}{|c|}{$\alpha '$ ($GeV^{-2}$ units)}\\ \hline
\multicolumn{2}{|c|}{0.228}&
\multicolumn{2}{|c|}{0.223}\\ 
\hline \hline
\multicolumn{4}{|c|}{$\alpha _I(0)$}\\ \hline
\multicolumn{2}{|c|}{-5.45}&
\multicolumn{2}{|c|}{-5.55}\\ 
\hline \hline 
\multicolumn{4}{|c|}{{\bf States of the Leading trajectories ($GeV$ units)}}\\ 
\hline 
{\it Predicted}&{\it Experimental}&{\it Predicted}&{\it Experimental}\\
\hline \hline
\multicolumn{4}{|c|}{J=1}\\ \hline
5.32 & 5.3248$\pm$.0018 & 5.41 &5.4163$\pm$.0033\\ 
\hline
\multicolumn{4}{|c|}{J=2}\\ \hline
5.72&/ &5.81&/ \\ 
\hline
\multicolumn{4}{|c|}{J=3}\\ \hline
6.09&/ &6.18&/ \\ 
\hline
\end{tabular}
\end{center}
\par
\vfill
\eject
\vskip1mm
\begin{figure}[htb]
\begin{center}
\epsfig{file=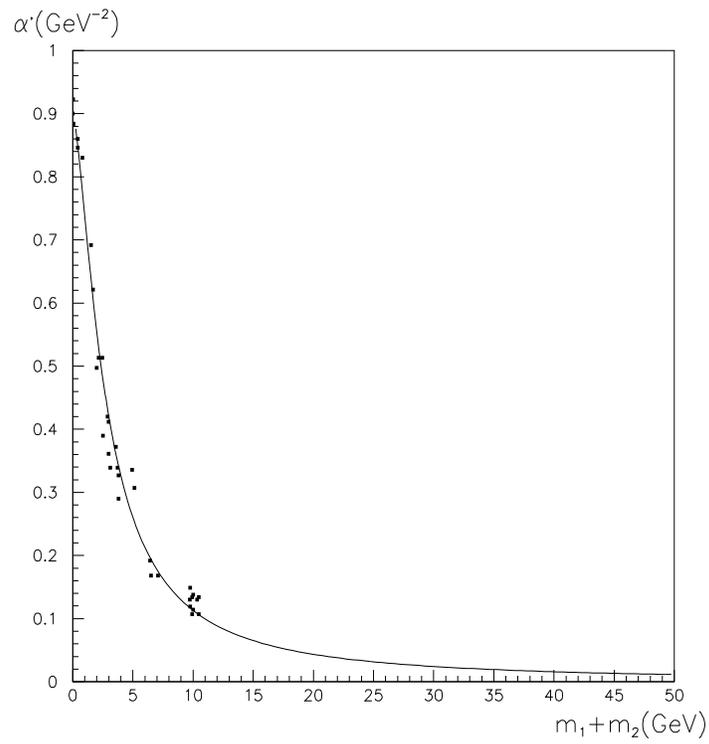,height=10.5cm,width=10cm}
\caption{{\em Slope parameter of the Regge trajectories as a function of
the sum of the constituent quark masses; input data compared with our 
analytic result Eq.3. }}
\end{center}
\end{figure}
\vskip1mm\noindent
\begin{figure}[htb]
\begin{center}
\epsfig{file=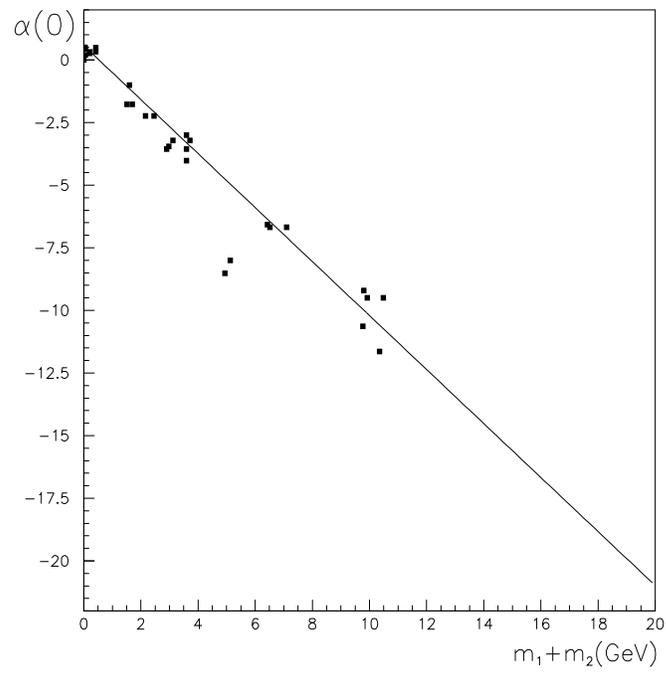,height=10cm,width=9.5cm}
\caption{{\em  The intercept parameter of the leading trajectory as a function
of the sum of the constituent quark masses; comparison of Eq.4 with input
data.}}
\end{center}
\end{figure}
\vskip1mm
\par\noindent
\vskip1mm
\begin{figure}[hbt]
\begin{center}
\epsfig{file=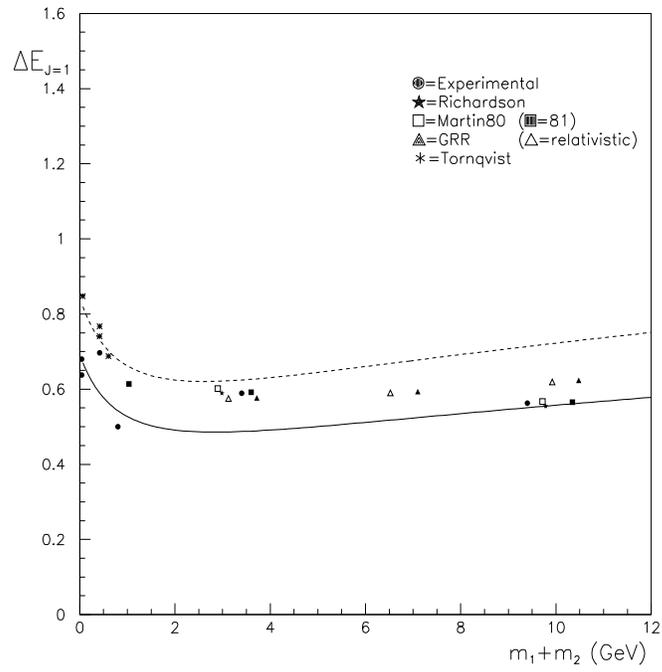,height=10cm,width=9.5cm}
\caption{{\em  Energy splitting for mesons of different flavors but same
 angular momentum; our predictions compared to available data and some 
 theoretical models.}}
\end{center}
\end{figure}
\vskip1mm
\vskip1mm
\begin{figure}[bht]
\begin{center}
\epsfig{file=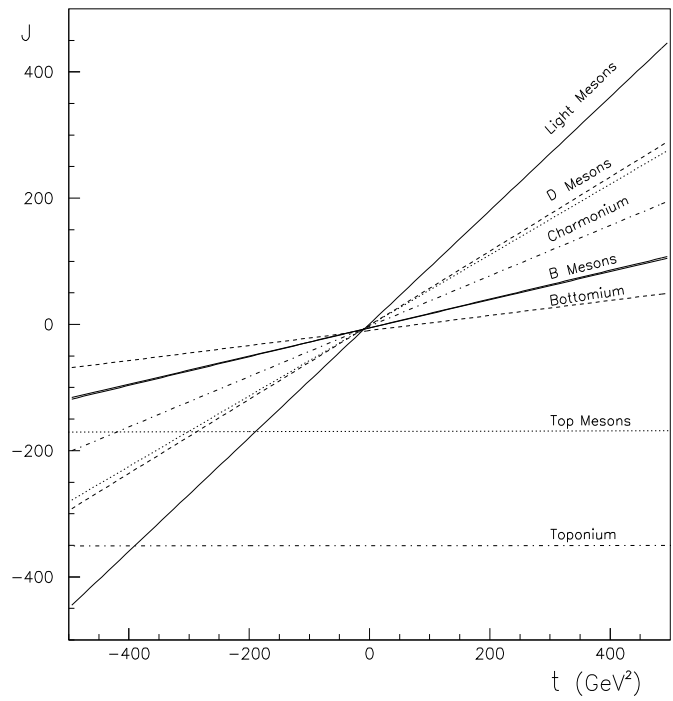,height=10.5cm,width=10cm}
\caption{{\em  Leading trajectories for different flavors for space- and
time- like regions, as given by Eq.(8).}}
\end{center}
\end{figure}
\end{document}